\documentclass[aps,twocolumn,pra,showpacs]{revtex4}%
\usepackage{graphicx}
\usepackage{amssymb}
\usepackage{color}
\usepackage{amsmath}
\usepackage{amsfonts}%
\begin{document}
\title{Ionization in a frozen Rydberg gas with attractive or repulsive potentials}
\author{Matthieu Viteau$^1$, Amodsen Chotia$^1$, Daniel Comparat$^1$, Duncan A. Tate$^*$$^1$, T. F. Gallagher$^\dagger$$^1$ and Pierre Pillet$^1$}
\affiliation{$^1$ Laboratoire Aim\'{e} Cotton,~CNRS, Univ Paris-Sud, B\^{a}t.~505, Campus d'Orsay, 91405 Orsay, France\\
$^*$ Department of Physics and Astronomy, Colby College, Waterville, Maine 04901-8858, USA\\
$^\dagger$ Department of Physics, University of Virginia, Charlottesville, Virginia 22904-0714, USA}

\date\today

\begin{abstract}
We report clear evidence of the role of dipole-dipole interaction in Penning ionization of Rydberg atoms, leading to the formation of an ultracold plasma. Penning ionization of np Rydberg Cesium atoms is prevented for states with in $n<42$, which correspond to a repulsive potential, but it does not occur for n larger than 42, corresponding to an attractive potential. Blackbody radiation is mostly responsible for the background and initial ionization, although ion-Rydberg collisions and population transfer due to limited superradiance may have to be considered.
\end{abstract}

\pacs{32.80.Ee; 34.20.Cf; 37.10.De}
\maketitle


A cold Rydberg gas is a fascinating system at the boundary of
atomic, solid state, and plasma physics. In a MOT, at $100\,\mu$K temperature, the atoms
move less than $~3\%$ of their typical separation, $10^{-3}\,$cm, on the
$1\,\mu$s time scale of the experiments, and such a sample resembles an
amorphous solid.  Since Rydberg atoms have large dipole moments,
scaling as the square of the principal quantum number $n$,
dipole-dipole interactions in the frozen Rydberg gas have a
significant effect, even though the atoms are far apart
\cite{Mourachko,Anderson,Fioretti}. For this reason, binary
dipole-dipole interactions have been proposed as the basis for
quantum gates \cite{Jaksch,Lukin}. Specifically, the dipole-dipole
interaction between a pair of Rydberg atoms can preclude the
excitation of the second atom in a sample once the first is excited,
a phenomenon termed the dipole blockade. Local, or partial,
blockades have been observed in many experiments, such as \cite{Tong} in Van der Waals case or \cite{VogtPRL2007} in dipole one. In addition to binary
interactions, there is evidence for many body interactions roughly
analogous to the diffusion of spins in  a glass
\cite{FrasierPRA99,Akulin99}. This phenomenon is particularly
apparent in the dipole-dipole energy transfer tuned into resonance
with an electric field, a process often termed the F\"orster
Resonance Energy Transfer (FRET) reaction \cite{VogtPRL2006}.

In addition a cold Rydberg gas can spontaneously evolve into a plasma.
If there is even a very slow ionization process, cold ions are
produced, and at some point their macroscopic space charge traps all
subsequent electrons produced \cite{Robinson,Killian,2007PhR...449...77K}. The trapped
electrons lead to a collisional avalanche which rapidly
redistributes the population initially put into a single Rydberg
state. Typically two thirds of the atoms are ionized and one third
are driven to lower states to provide the requisite energy
\cite{Robinson,Walz}. The origin of the initial ions is not,
however, completely understood \cite{Amthor,Day2008,ReinhardPRL2008}.  With a pulsed laser it is possible
to excite atoms close enough to each other that they interact
strongly, resulting in ionization on a $100\,$ns time
scale, too fast to be a result of motion of the atomic nuclei
\cite{LiPRA2004}.

With narrow bandwidth, quasi continuous wave (cw) excitation, it is
not possible, due to the energy shift produced by the dipole-dipole interaction, to excite atoms which are close together, yet ionization still
occurs, although on a time scale of microseconds \cite{Tanner}. One
mechanism for this ionization is that pairs of atoms excited to attractive
diatomic potential curves collide, resulting in
the ionization of one of the atoms. The much more rapid conversion
of cold Rb $nd$ atoms to a plasma than $ns$ atoms was attributed by Li
$et$ $al.$ to ionizing collisions of pairs of atoms excited to attractive
potentials in the $nd$ case but not in the $ns$ case \cite{LiPRL2005}.
Using a narrow bandwidth laser Vogt $et$ $al.$ showed that ions are present on attractive Cs $npnp$ potentials \cite{VogtPRL2006}. 
A similar behaviour was noticed by Amthor $et$ $al.$ where they demonstrated that
pairs of atoms excited closer together on the Rb $60d60d$ attractive potential
ionized more rapidly than those farther apart. A surprising aspect of their observations is the high ionization rate
for the $62s$ state in spite of the repulsive $62s62s$ potential \cite{Amthor}.

In an effort to isolate the effect of attractive/repulsive
potentials from other effects, we have examined the ionization of
cold Cs $np$ atoms excited with narrow bandwidth excitation. The interest
of the Cs $np$ states is that for $n>42$ a pair of Cs $np$ atoms is on
an attractive potential, while for $n<42$ the potential is
repulsive. Thus by varying $n$ from $40$ to $45$ we switch, at $n=42$,
from excitation to a repulsive potential to an attractive one, with
all other parameters of the system changing only slowly. The
results show unambiguously the difference between attractive and
repulsive potentials. Perhaps as interesting, in the repulsive case
we observe ionization which is nearly linear in the number of
excited atoms, but at a rate that is twice that due to blackbody photoionization. Our
observation of such a high ionization rate is not unique; others have
made similar observations \cite{Amthor}. We suggest here that possible sources of high ionization
rates are ionizing collisions, due to motion on attractive
ion-dipole potentials and dipole-dipole potentials activated by state changes stimulated by the blackbody radiation and superradiant decay. In the sections which follow we describe the Cs
system we have studied, present our experimental results, and
compare them to expectations based on $300$K black body population
transfer.

\begin{figure}[h!]
    \centering
        \resizebox{0.4\textwidth}{!}{
        \includegraphics*[0mm,0mm][89mm,134mm]{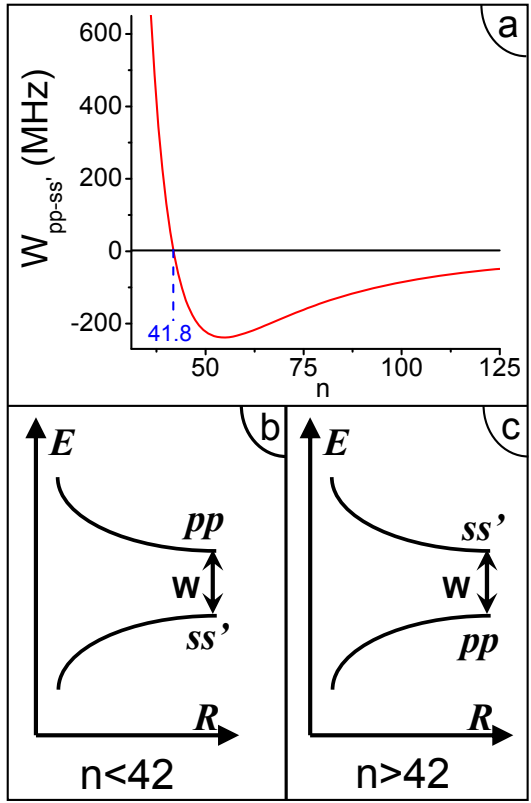}}
        \caption{(a) The energy difference (W) between the
molecular pp and ss' states at $R=\infty$ for different n. (b) and (c) Schematic potential curves of pair of atoms as function of internuclear distance. }
    \label{fig:Fig1}
\end{figure}

For convenience we consider the molecular state composed of a pair
of Cs Rydberg atoms separated by distance R. A pair of $np$ atoms we
term the $pp$ state, and the nearly degenerate pair of $ns$ and $(n+1)s$
atoms we term the $ss'$ state. The energy difference between the
molecular $pp$ and $ss'$ states at $R=\infty$ is given by
\begin{equation}
W_{pp-ss'}=2W_{np}-W_{ns}-W_{(n+1)s}
\label{equ1}
\end{equation}
where $W_{nl}$ is the energy of the Cs $nl$ state. As shown by
Fig.\ref{fig:Fig1}a, $W_{pp,ss'}$ crosses zero at $n\sim42$.  For
$n>42$ the $pp$ state lies below the $ss'$ state, while for $n<42$ the
reverse is true. If we ignore the fine-structure and the angular distribution, the $pp$ and $ss'$ states are coupled by the
dipole-dipole interaction $\frac{\mu\mu'}{R^3}$, where $\mu$ and
$\mu'$ are the dipole matrix elements connecting the $np$ state to
the $ns$ and $(n+1)s$ states, and at finite $R$ the eigenstates are
attractive and repulsive linear superpositions of $pp$ and $ss'$. The resulting potential curves are schematically shown in
Fig.\ref{fig:Fig1}b and Fig.\ref{fig:Fig1}c. As $R\rightarrow\infty$
the potentials are $1/R^6$ van der Waals potentials, but at small
$R$ they are $1/R^3$ dipole-dipole potentials \footnote{\label{pot}The ionization time is simply calculated by integrating the  dipole-dipole forces between two Rydberg atoms: $U(R)=\frac{W_{pp-ss'}}{2}\left(1+\sqrt{1+(\frac{\mu\mu'}{R^3W_{pp-ss'}/2})^2}\right)$}. We excite Cs atoms to
the atomic $np_{3/2}$ state, or pairs of atoms to the molecular $pp$
state. The repulsive potential is excited for $n<42$, the attractive
potential for $n>42$, while for $n=42$ at high atomic density, both potentials are excited.

In the experiment Cs atoms are held in a magneto optical trap (MOT)
at a temperature of $100\,\mu$K and a number density of up to
$5\times10^{10}\,$cm$^{-3}$.  The atoms are excited to the $np_{3/2}$ states via
the route
\begin{equation}
    6s_{1/2} \rightarrow 6p_{3/2} \rightarrow 7s_{1/2} \rightarrow np_{3/2}
    \label{equ2}
\end{equation}
using three cw single frequency lasers. The first laser is the
$852$nm trap laser, a diode laser which is typically left on
continuously. Using acousto optic modulators we form the outputs of
the second and third lasers into temporally overlapping pulses at an
$80$Hz repetition rate. The second laser is a $1470$nm diode laser
with a typical power of $20$mW.  It is focused to a beam waist of
$100\mu$m.  The third laser is a Titanium:Sapphire laser operating
near $830$nm. It has a beam waist of $70\mu$m, and it crosses the
second laser at $67.5^{\circ}$, producing a $2\times10^{-3}\,$mm$^{3}$ volume
of Rydberg atoms. The second and third laser pulses have a 300$\,$ns durations.  The inherent limitation of the linewidth of
the final transition is imposed by the $54\,n$s lifetime of the Cs
$7s$ state, the $1\,$MHz laser linewidth and the Fourier transform of the pulse duration.  Subsequent to its
production by the laser pulses the cold Rydberg gas is allowed to
evolve for times from $450\,$ns to $50\,\mu$s. At this time we analyze
the population with a field ionization pulse which rises to
$500\,$V/cm in $50\,$ns, applied using two grids $1.57\,$cm apart.  Any
ions present are ejected from the MOT at the beginning of the field
ionization pulse, the  Rydberg atoms are ionized, and the
resulting ions are ejected later.  Both sets of ions strike a
microchannel plate detector, producing time resolved ion and Rydberg
atom signals for initial states of $n<80$, which are registered with
two gated integrators and stored in a computer.

\begin{figure}[h!]
    \centering
        \resizebox{0.8\textwidth}{!}{
        \includegraphics*[12mm,136mm][207mm,271mm]{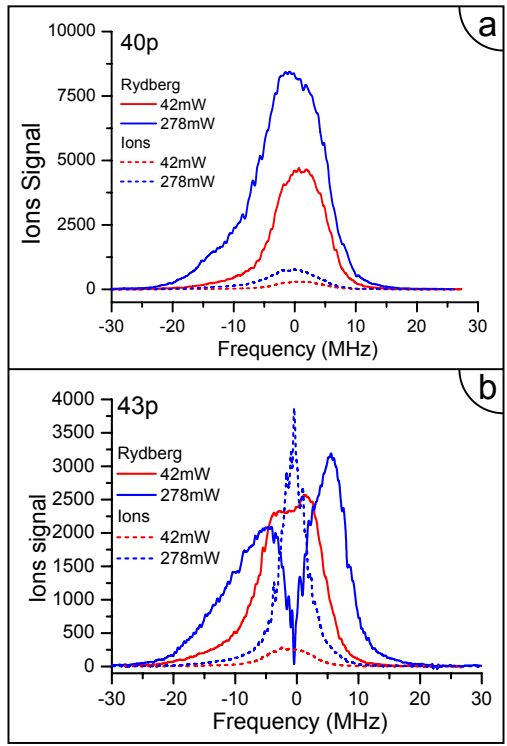}}
        \caption{Rydberg and Ion signal with two different laser intensity (42mW and 278mW) for (a) $40p$ state (repulsive) and (b) $43p$ (attractive) Data are taken after a time delay  of $10\,\mu$s,
between the end of the laser pulse and the field ionization pulse. }
    \label{fig:Fig2}
\end{figure}

The data are taken by scanning the frequency of the third laser
while recording the ion and atom signals, with all other parameters
held fixed.  In Fig.\ref{fig:Fig2}a and Fig.\ref{fig:Fig2}b we show
typical recordings of the ion and atom signals for the $40p_{3/2}$
and $43p_{3/2}$ states respectively, for two different laser
intensities. We note the asymmetrical shoulder on the red side of the line, due to
the three step excitation, where direct two and three photon
excitations of process (\ref{equ2}) can occur. The time delay between the end of the laser pulse and the field ionization pulse is
$10\,\mu$s. At low power, i.e. at low Rydberg density, we observe in both
cases a small number of ions, ($7\%$ of Rydberg atoms excited for $40p$ and $10\%$ for $43p$). When we increase
the density, for $40p$ the number of ions is still small ($9\%$) but
for $43p$ we observe, at resonance, complete ionization of the Rydberg sample, the
signature of the formation of a plasma. The differences between
$n=40$ and $n=43$ in the frequency and magnitude of the ionic
signals are due to the excitation of atoms to the attractive
potential in the latter case, as suggested by \cite{LiPRL2005}. 

\begin{figure}[h!]
  \centering
  \resizebox{0.4\textwidth}{!}{
    \includegraphics*[0mm,0mm][45mm,97mm]{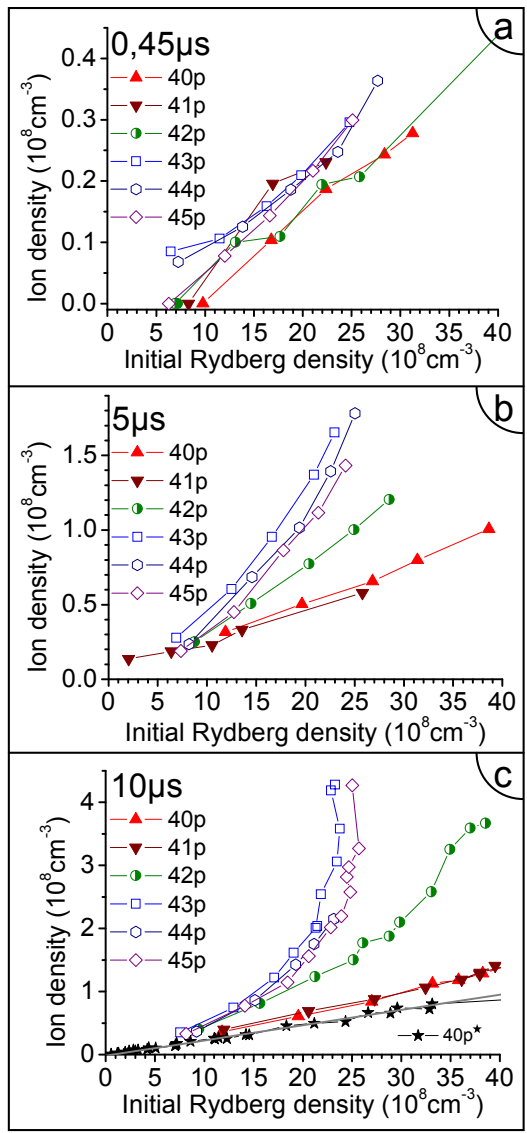}}
  \caption{Ion density as function of initial Rydberg density for different $n$ states and three delay time (a) $0.45$ (b) $5$ and (c) $10\,\mu$s. At $10\,\mu$s the ``40p$^{\star}$" is without $6p$ cold and hot atoms. Each point corresponds to the maxima of spectral scans, such as those shown in
Fig.\ref{fig:Fig2}.}
    \label{fig:Fig3}
\end{figure}

The maxima of spectral scans, at a fixed frequency, such as those shown in
Fig.\ref{fig:Fig2} for a range of densities, delay times and quantum
states give a more comprehensive picture. In Fig.\ref{fig:Fig3}
we show the density of ions vs density of Rydberg atoms initially
excited, for $0.45$, $5$, and $10\,\mu$s delay times. The number of Rydberg atoms was
varied by changing the intensity of the Ti:Sapphire laser. We note that there is a large number of ions present at a
delay of $0.45\,\mu$s, and we believe they are formed during the $300\,$ns laser
excitation. The central features of Fig.\ref{fig:Fig3} are the following.  With a
$0.45\,\mu$s delay the ionization yields for all n states are
essentially the same, but with $5$ and $10\,\mu$s delays there is a
clear difference between the $n<42$ and $n>42$ states. The only
difference between these two cases is that the atoms are excited to
a repulsive potential in the former case and an attractive one in
the latter case. Two atoms excited to the $43p43p$ state with a
separation of $5\,\mu$m collide and ionize in $6\,\mu$s \ref{pot}[24].
For a Rydberg density of $25\times10^{8}\,$cm$^{-3}$, corresponding to $5000$ Rydberg
atoms in the trap, $10\%$ of the atoms are this close together, so
the difference between the $n<42$ and $n>42$ behaviours is evidently due
to the dipole-dipole induced collisions.

For the $10\,\mu$s delay the ion production, for $n>42$, starts to become very
nonlinear for a ion density of $~2\times10^8\,$cm$^{-3}$ ($\sim400$ ions), which we
attribute to the trapping of the electrons by the ions and the
subsequent ionizing collisions of the electrons with the Rydberg
atoms. 

For $5$ and $10\mu$s delays, the $42p$ state is between the two sets of curves, corresponding to an excitation to both the repulsive and attractive potentials.
While the difference between the attractive and repulsive curves of Fig.\ref{fig:Fig3}c is understood, the source of the ionization for the $n<42$
states is less evident.

 We show two kinds of
data for the repulsive case ($n<42$). For the $40p$ and $41p$ data, $6p$ cold and hot
atoms are present due to the trap lasers. For the data labelled ``40p$^{\star}$" the
$6s-6p$ excitation is turned off at the same time as the second and
third lasers are turned off, so there are no $6p$ atoms with which to
interact. In the case of the $40p$ state with no $6p$
atoms, we observe an apparently density
independent ionization rate of $2400\,$s$^{-1}$. This linear rate also appears to
be present in the $n > 42$ states at low density. The calculated blackbody photoionization rate
is $1200\,$s$^{-1}$ \cite{LiPRL2005, 2007PhRvA..75e2720B}. The observed rate is twice the calculated rate.  Our observations are not unique in
this regard \cite{Amthor}. The ionization rates for excitation to repulsive curves
are generally higher than the calculated blackbody photoionization
rates, and we now consider possible causes of this ionization.

One possibility is the black-body transfer of population from the $np$ states to the nearby $ns$ and $nd$ states, producing $ns-np$ and $nd-np$ pairs, with a rate $\sim10\,000$s$^{-1}$. Roughly half the pairs will be on attractive $1/R^3$ potentials and half on the repulsive potentials. Only $5\%$ of those on attractive potentials can collisionally ionize on a 10$\,\mu$s time scale, corresponding to a rate of ions formation around $250$s$^{-1}$, which is not fast enough to explain our observations. This rate could be, for short time scales, increased by superadiance \cite{WangPRA2007,Day2008}. In this connection we note that, for some conditions (medium density), the Cs $40p$ state could exhibits a rapid initial decay, with a $\sim10\,\mu$s decay time, far faster than the 300$\,$K $40p$ lifetime of $47\,\mu$s. But in most of the conditions of this article the presence of the dipolar and ionic dephasing would probably limit this effect.

We suggest here a new contribution to the large ionization rates in the
case of the repulsive potentials based on the effect of cold ions, which may attract the nearest Rydberg atom by the ion-dipole interaction. The process could be the following: one ion and one Rydberg could collide to produce a translationally cold Cs$^+_2$ molecule, which in turn attracts another Rydberg atom, a process likely to result in two ions. However, the ions are hot, and the process terminates, so this could increase the ionization rate up to $1000$s$^{-1}$ rate. Taken together the above mechanisms are probably responsible for the observed ionization rates for the $n<42$ states. We note, though, that the scenarios outlined above should lead to a nonlinear ionization rate for short times or low densities, which is not apparent in our data, although the fact that the sample of atoms does not have a uniform density may mask such features. The notion of collisions due to ion-Rydberg dipole attractive potentials assumes that the
polarizability, of the Rydberg atom, is greater than zero,
that is the Stark shift is to lower energy when the electric field is increased. This requirement is met
for the Cs $40p$ and Rb $40s$ states, but not for the Cs $39d$ state,
and it may be why the ionization rate $\sim1000$ s$^{-1}$ of the Cs $39d$ state  \cite{Bruno} is close
to the calculated blackbody photoionization, but the ionization
rates of the Cs $40p$ and Rb $40s$ states are not. 

In summary, the measurements reported here demonstrate clearly that
excitation to an attractive, as opposed to a repulsive potential
dramatically increases the initial ionization rate due to
dipole-dipole induced collisions. However, atoms excited to
repulsive curves also ionize at rates in excess of the expected blackbody
ionization rates. We suggest that in this case the attractive
atom-atom or ion-atom potentials lead to increased ionization. The complete dynamics of the ionization of an ensemble of Rydberg atoms is complex, but the attractive dipole-dipole or dipole-ion forces are the main parameter for this problematic.

The authors acknowledge fruitful discussions with Vladimir Akulin. This work is in the frame of ``Institut Francilien de
Recherche sur les Atomes Froids" (IFRAF)


\end{document}